\documentclass[%
 reprint,
superscriptaddress,
nofootinbib,
 amsmath,amssymb,
 aps,
 prr,
]{revtex4-2}

\usepackage{siunitx}
\usepackage{graphicx}
\usepackage{dcolumn}
\usepackage{bm}
\usepackage{hyperref}
\usepackage{mathtools}
\usepackage{pifont}

\usepackage[dvipsnames]{xcolor}

\newcommand{\im}{\mathrm{i}}

\newcommand*\de{\mathop{}\!\mathrm{d}}

\newcommand{\Oh}{\mathcal{O}}

\renewcommand*{\Re}{\operatorname{Re}}
\renewcommand*{\Im}{\operatorname{Im}}

\newcommand{\myeqref}[2]{(\hyperref[#1]{\ref*{#1}#2})}
\newcommand{\myref}[2]{\hyperref[#1]{\ref*{#1}(#2)}}
\newcommand{\myrefnb}[2]{\hyperref[#1]{\ref*{#1}#2}}
\newcommand{\subtag}[1]{\tag{\theparentequation #1}}

\newcommand*{\ie}{\emph{i.e.}}
\newcommand*{\eg}{\emph{e.g.}}

\newcommand{\w}{w}

\newcommand{\n}{\hat{\nu}}
\newcommand{\s}{\hat{s}}

\usepackage[normalem]{ulem}

\newcommand{\Er}{\mathrm{Er}}
\newcommand{\A}{\mathit{A}}
\newcommand{\M}{\mathit{M}}

\newcommand{\I}{\bm{\mathrm{I}}}
\newcommand{\E}{\bm{\mathrm{E}}}
\newcommand{\W}{\bm{\Omega}}

\newcommand{\scdot}{{\mspace{1.5mu}\cdot\mspace{1.5mu}}}
\DeclareMathOperator\arctanh{arctanh}

\begin{document}


\title{Active nematic response to a deformable body or boundary:\\ elastic deformations and anchoring-induced flow}

\author{Thomas G.~J.~Chandler}
\email{tgchandler@wisc.edu}
\author{Saverio E.~Spagnolie}%
\email{spagnolie@wisc.edu}
\affiliation{%
Department of Mathematics, University of Wisconsin–Madison, Madison, WI 53706, USA}%

\date{\today}

\begin{abstract}
A body immersed in a nematic liquid crystal disturbs the fluid's preferred molecular configuration and increases its stored elastic energy. In an active nematic, the fluid components also generate a stress in the bulk fluid. By introducing either an immersed body or boundary, a large scale flow can be triggered due to anchoring boundary conditions alone --- a global pressure built by active stresses at equilibrium is instantly released everywhere. The fluid then imposes viscous, elastic, and active stresses on such surfaces which, if compliant, may result in a surface deformation. We study the deformations and stresses of a linearly elastic body placed in an active nematic in two dimensions. Using complex variables techniques, exact expressions for the fluid flow, director field, surface tractions, and  body deformation are derived. Qualitative differences between elastic and active stress-driven deformations are identified, depending on an active Ericksen number, anchoring conditions, and body material properties, thereby suggesting a new method for measuring mechanical stresses in active anisotropic environments. Flow profiles, external confinement, and anchoring-induced stirring are also addressed.
\end{abstract}


\maketitle

\section{Introduction}

The eukaryotic cell is host to a wide variety of deformable organelles, from the nucleus to mitochondria \cite{il13,hmechhsp16,uk17,chz17,rohk19,kslg22}. These organelles reside in complex environments which can be anisotropic \cite{dnllc08,glg19}, and experience mechanical stresses due to actively driven cytoplasmic flows \cite{hw93,kddsg15,pjj15,mm18,dpkndpvdd19,hl23,dflkbslgss24}. The cell membrane itself is also deformable, and its morphology is commonly influenced by internal processes. For instance, active anisotropic stresses are of functional importance in cytoskeletal remodeling of the nucleus and chromatin configurations \cite{rs15,ssz18,sbm19,Zidovska20,wcrkgacadl21}, in the metaphase spindle \cite{Shelley16,onb18,ojb20}, and in numerous other aspects of tissue morphogenesis \cite{afssrje10,molg15,bbc24}. On a larger length scale, active systems like bacterial swarms and biofilms also exhibit anisotropy, among other complex rheological properties \cite{sbspw08,fw10,sa12,Saintillan18,su23}. 

Even when free of bodies, actively-driven anisotropic flows exhibit very rich and complex dynamics \cite{nsml97,dccgk04,ss08,scdhd12,diys18}. Stability \cite{sr02,ss08,sk09,Ramaswamy10,hs10,gmch12,tgy14,Weady24}, topology \cite{Giomi15}, pattern formation \cite{gmch11,dhdwbg13,os22}, and mixing \cite{trsoafmh19,mhsemh24} have seen substantial theoretical treatment. Confinement introduces an additional length scale, which can affect the emergent spatial structure \cite{tas17,jznfbh23}. Strong active nematic stresses can also change the shape of deformable confining surfaces  \cite{klsdgbmdb14,bn14, zzrd16,lmbm17, slpp19,Alert22,lgab24,fs24}, and can play a prominent role in directing the motility of cells \cite{mo96,Svitkina18} and active droplets \cite{mkhb16,gl17, wzmda21,yss21}. 

When an active nematic contains an immersed body, or inclusion, additional elastic stresses are immediately introduced. In passive systems, these stresses result in intricate and beautiful textures \cite{Stark01,Musevic17,Smalyukh18}, and can be used to measure cell material properties  \cite{nesa20,ghahan24}. In active systems, fluid elasticity and anisotropy can redirect the emergent flows, leading to complex body dynamics. Recent examples include an investigation of the biased rotation of gear-like bodies \cite{rzd23,ha23}, complex droplet trajectories \cite{sc24,nhcsmgt25}, and transitions from fixed point to limit cycle to chaotic dynamics of a circular disk \cite{Freund23}. How body deformability affects and is affected by such an active environment remains largely unexplored. 



In this paper, we study an active nematic fluid based on Ericksen–Leslie theory \cite{Landau1986, degennes1993,stewart2004,mjrlprs13,diys18} with a soft internal or external boundary in two dimensions. Complex variable techniques yield exact analytical representations for the director field, velocity field, and boundary deformation. Elastic deformations of bounding surfaces are known to  relieve the elastic energy stored in the bulk LC phase \cite{cs24a}. Here, we show that active stresses can compete and eventually dominate this balance, resulting in distinct geometric signals. In addition, we show that a spontaneous stirring flow with a unique plume structure may be induced with a change in surface anchoring boundary conditions alone, due to a global pressure release.

\section{Active Nematic Flows}

Active nematic liquid crystals (LCs) can be described by their locally-averaged molecular orientation (\ie~the director field) $\bm{n}(\bm{x},t)$ and fluid velocity $\bm{u}(\bm{x},t)$, with spatial position $\bm{x}$, time $t$, and $|\bm{n}|=1$. The director field, $\bm{n}$, evolves due to gradients in the velocity field, with strain-rate tensor $E_{ij}=(\partial_iu_j+\partial_ju_i)/2$ and vorticity tensor $\Omega_{ij}=(\partial_iu_j-\partial_ju_i)/2$, and a relaxation towards equilibrium under the action of the  molecular field, $\bm{H}=-\delta \mathcal{F}/\delta \bm{n}$, where $\mathcal{F}$ is the  free energy density. In the one-constant approximation, $\mathcal{F}=K\lVert\nabla\bm{n}\rVert^2/2$ and so $\bm{H}=K\nabla^2\bm{n}$, for the single (Frank) elastic constant $K$ \cite{degennes1993}. Together,
\begin{equation}\label{eq:directoreq}
(\partial_t+\bm{u}\cdot \nabla) \bm{n} =\bm{n}\cdot\W+ \lambda\bm{n}\cdot \E\cdot(\I-\bm{n}\bm{n})+\bm{h}/\gamma,
\end{equation}
with $\gamma$ the rotational viscosity, $\lambda$ the tumbling/reactive parameter, and $\bm{h}=(\I-\bm{n}\bm{n})\cdot \bm{H}$ the transverse part of the molecular field, where $\bm{n}\bm{n}$ is a dyadic product \cite{Landau1986,degennes1993,stewart2004,cs24a}. Momentum and mass conservation, under the assumptions of incompressibility and zero Reynolds number flow\footnote{The Reynolds number is defined as $\Re=\rho |\alpha| L^2/\mu^2$, with $\rho$ the fluid density, and other terms defined above.}, are given by
\begin{subequations}\label{eq:conservation}
\begin{equation}
\nabla\cdot\left(\bm\sigma^e+\bm\sigma^v+\bm\sigma^a-p \I\right)=\bm{0}\quad
   \text{and}\quad \nabla\cdot\bm{u}=0,\subtag{a,b}
   \end{equation}
\end{subequations}
respectively, for $\nabla\cdot\bm{\sigma}\coloneqq\partial_{i}\sigma_{ij}$,  pressure $p$, and  elastic, viscous, and active stress tensors
\begin{subequations}\label{eq:stresstensors}
\begin{align}
\begin{split}
   \bm \sigma^e &=\frac{K}{2}\lVert\nabla\bm{n}\rVert^2\mathbf{I} -K\nabla \bm{n}\cdot \nabla \bm{n}^T \\
   &\quad-\frac{1}{2}(\bm{n}\bm{h}-\bm{h}\bm{n})-\frac{\lambda}{2}(\bm{n}\bm{h}+\bm{h}\bm{n}),
   \end{split}
   \\
    \bm \sigma^v &= 2\mu \E+\mu_1(\bm{n}\scdot\E\scdot\bm{n}) \bm{n}\bm{n}+\mu_2 (\bm{n}\E\scdot\bm{n}+\bm{n}\scdot\E\bm{n}),\\
    \bm  \sigma^a &= 2\alpha \bm{n}\bm{n},
\end{align}
\end{subequations}
respectively. Here, $\mu$ is the solvent viscosity, $\mu_1$ and $\mu_2$ are anistropic viscosities, and $\alpha$ is the activity strength, which can describe both extensile ($\alpha<0$) and contractile ($\alpha>0$) activity \cite{mjrlprs13,ss15,diys18} (see \cite{SuppMat}). Note the convention taken for the divergence, which is important since $\bm{\sigma}^e$ is not symmetric.

The system is made dimensionless by scaling upon an intrinsic length scale $L$ (\eg~the size of an immersed body), velocity scale $U$ (to be defined), and stress scale $K/L^2$. The system is then characterized by the dimensionless Ericksen number, activity strength, and  viscosity ratios \cite{SuppMat}:
\begin{subequations}\label{eq:Pars}
\begin{equation}
    \Er\coloneqq\frac{\mu U L}{K},\quad \A \coloneqq\frac{\alpha L}{\mu U},\quad \gamma' \coloneqq \frac{\gamma}{\mu},\quad \mu_j'\coloneqq\frac{\mu_j}{\mu}.\subtag{a--d}
\end{equation}
\end{subequations}

For small rotational Ericksen numbers, $\gamma'\Er\ll 1$, the fluid flow does not affect the director field, and at leading order from \eqref{eq:directoreq} we have
\begin{equation}
    \bm{h}=(\mathbf{I}-\bm{n}\bm{n})\cdot\nabla^2\bm{n}=\bm{0}.
\end{equation}
Further, assuming small viscosity ratios, $\gamma',\mu_i'\ll1$, Eq.~\myeqref{eq:conservation}{a} yields
\begin{equation}\label{eq:momenteq_smallEr}
    -\nabla p+\nabla^2\bm{u}+2\A \nabla\cdot(\bm{n}\bm{n})=\bm{0}.
\end{equation}

In the examples presented in this paper, the fluid flow is induced by activity alone; thus, the velocity scale is set by the activity strength, $U=|\alpha|L/\mu$, and the  assumption of a small  Ericksen number is equivalent to assuming weak activity, $\gamma'\Er =|\alpha|\gamma L^2/(\mu K)\ll1$. \footnote{In 2D incompressible flows, a general $\mu_2$ can actually be absorbed into the isotropic  viscosity by taking $\mu\mapsto \mu-\mu_2/2$.}

\begin{figure}[htp!]
    \centering
    \includegraphics[width=\linewidth]{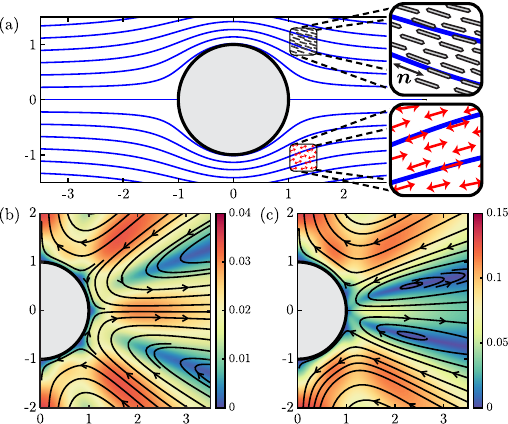}
    \caption{The activity-induced flow of a 2D nematic LC outside a  cylinder for  $\A=-1$ (\ie~extensile activity). (a) The director lines for $w=10$  are shown as blue curves. (b,c) The activity--induced velocity field is shown for (b)~$w=1$ and (c)~$w=10$. The black arrows show the flow direction and the color denotes the fluid speed, $\sqrt{u^2+ v^2}$. Contractile activity with $\A=1$ corresponds to flipping the flow direction.}
    \label{fig:Fig1}
\end{figure}

\subsection{Two-dimensional active nematics}

In 2D, the director field can be described by an angle field $\theta(x,y)\in[0,\pi)$ such that $\bm{n}=(\cos\theta,\sin\theta)$, and a streamfunction $\psi(x,y)$ can be introduced such that $\bm{u}=(\partial_y\psi,-\partial_x\psi)$. The director field, governed by $\bm{h}=\bm{0}$, reduces to Laplace's equation for $\theta$, while the momentum equation \eqref{eq:momenteq_smallEr} is equivalent to Poisson's equation for the scalar vorticity, $\omega=-\nabla^2\psi$:
 \begin{subequations}\label{eq:asym_sym2d}
 \begin{gather}
 \qquad\nabla^2\theta=0,\\
 \nabla^2\omega=\A \left[2\partial_{yx}\cos(2\theta)+(\partial_{yy}-\partial_{xx})\sin(2\theta)\right]. \end{gather}
 \end{subequations}
Here, we  absorb the isotropic parts of the stress tensors into the pressure by taking $p\mapsto p+\A$. The leading-order, dimensionless, traceless stress tensors are  denoted with hats:
\begin{subequations}\label{eq:tracelessstress}
 \begin{align}
   \bm{\hat\sigma}^e &= \frac{1}{2}|\nabla \theta|^2\mathbf{I} -\nabla \theta\nabla \theta,\\
     \bm{\hat\sigma}^v &= \nabla\bm{u}+\nabla\bm{u}^T,\\
     \bm{\hat\sigma}^a &= 2 \bm{n}\bm{n}-\mathbf{I},
 \end{align}
 \end{subequations}
giving the dimensionless (bulk)   stress tensor  $\bm{\sigma}=\bm{\hat\sigma}^e+\Er\left(\bm{\hat\sigma}^v+A\bm{\hat\sigma}^a-p\mathbf{I}\right)$.

At any boundaries, we assume  the fluid velocity vanishes  and the director field relaxes locally towards a preferred angle $\theta_0(s)$ due to surface anchoring of strength $W$ (\ie~Rapini--Papoular anchoring~\cite{rp69}):
\begin{subequations}\label{eq:2D-bcs}
\begin{equation}
\partial_s\psi=\partial_\nu\psi= 0 \quad \text{and}\quad \partial_{\nu}\theta=\frac{w}{2}\sin\left[2(\theta-\theta_0)\right], \subtag{a,b}
\end{equation}
\end{subequations}
for the anticlockwise arclength derivative $\partial_s=\bm{\s}\cdot\nabla$, normal derivative pointing into the LC $\partial_{\nu}=\bm{\n}\cdot\nabla$, and the dimensionless anchoring strength $w\coloneqq WL/K.$

With $z=x+\im y$ and $\bar{z}=x-\im y$ its complex conjugate, the system may be written in complex variables as
\begin{subequations}\label{eq:complexeqs}
\begin{equation}
\partial_{z\bar{z}}\theta=0\quad\text{and}\quad
\partial_{\bar{z}}(p-\im\omega)=\A \partial_{z}e^{2\im\theta},\subtag{a,b}
\end{equation}
\end{subequations}
with vorticity $\omega = - 4\partial_{z\bar{z}}\psi$ and velocity, $\bm{u}=(u,v)$, given by $u-\im v  = 2\im \partial_z\psi$.  Integrating  \eqref{eq:complexeqs} yields 
\begin{subequations}\label{eq:thpom}
\begin{align}
\theta&= -\arg f'(z) \\
\text{and}\quad p-\im\omega &=\A \partial_z\bigl[g(z)+ \overline{f(z)}/f'(z)\bigr],
\end{align} 
\end{subequations}
for some locally-holomorphic functions $f$ and $g$ \cite{ablowitz2003}. We  assume the LC does not contain any topological defects, so $f$ is non-zero and singularity free inside the LC, though this assumption can be relaxed \cite{mn2022,ms2024}.
Inserting the imaginary part of \myeqref{eq:thpom}{b} into  $ \partial_{z\bar{z}}\psi=-\omega/4$ and integrating yields
\begin{equation}\label{eq:psi}
\psi =\frac{\A }{4}\Im\Bigl[\bar{z}g(z)+h(z)+\overline{\int f(z)\de z}\,/f'(z)\Bigr],
\end{equation} 
for another  locally-holomorphic function $h$.  The first two terms in \eqref{eq:psi} correspond to the well-known form of a biharmonic function with Goursat functions $h$ and $g$ \cite{Langlois1964}, while the last term accounts for the active forcing.

Analytical solutions for the director and velocity fields are  obtained, provided that functions $f$, $g$, and $h$, which are locally holomorphic within the LC and satisfy the boundary conditions in \eqref{eq:2D-bcs}, may be found. The  problem for $f$ derived here  is equivalent to the potential problem explored in Refs.~\cite{cs23,cs24} for a passive LC, but $g$ and $h$ depend on activity.  

\subsection{Anchoring-controlled flows}

As an example, consider a cylinder of radius $L$ of infinite extent immersed in a weakly-active nematic LC. In the far-field, we assume the LC is aligned horizontally in a direction orthogonal to the cylinder's long axis and the fluid velocity vanishes. The LC is assumed to be  tangentially anchored to the body with finite strength, \ie~\myeqref{eq:2D-bcs}{b} with $\theta_0=\arg (\im z)\mod\pi$  on $|z|=1$. Without the body, the system sits at an equilibrium, with $\bm{n}=\bm{\hat{x}}$ and $\bm{u}=\bm{0}$, and the active stress is absorbed by the pressure.

To find the activity-induced flow, we first determine the LC director angle $\theta=-\arg f'$. Here, the potential $f(z)$ must be locally holomorphic in $1<|z|<\infty$ and satisfy the finite anchoring condition  on $|z|=1$, and $f(z)\sim z$ as $|z|\to\infty$. The solution to this problem is
\begin{equation}\label{eq:cylinder_theta}
\theta =-\arg f'(z)= -\arg(1-\rho^2/z^2),
\end{equation}
where $\rho(w) \coloneqq({\sqrt{1+4/\w^2}-2/\w})^{1/2}$ is an effective  radius \cite{cs23}. The corresponding director lines  (which lay parallel to $\bm{n}$) are plotted in Fig.~\myref{fig:Fig1}{a} for $w=10$. Meanwhile, the Goursat functions $h(z)$ and $g(z)$ must be locally holomorphic in $1<|z|<\infty$ and yield a fluid velocity that vanishes on $|z|=1$ and as $|z|\to\infty$. This ultimately yields the streamfunction (see \cite{SuppMat}):
\begin{equation}\label{eq:cylinder_psi}
    \psi= \frac{\A\rho^2}{8}\Im\biggl[\frac{(|z|^2-1)^2}{z^2(z^2-\rho^2)}+\frac{1-|z|^2+2\log |z|}{(z^2-\rho^2)/(2\rho^2)}\biggr].
\end{equation}
The corresponding flow is shown in Fig.~\myref{fig:Fig1}{b,c} for $w=1$ and $w=10$, with $\A=-1$ (\ie~extensile activity).  

A prominent feature of the induced flow is a plume extending symmetrically to both the left and right of the cylinder (see Fig.~\myref{fig:Fig1}{b,c}). Far from the body, 
\begin{equation}
    u+\im v =-2\im\partial_{\bar{z}}\psi \sim \bigl[\rho^2\cos(2\vartheta)-\cos(4\vartheta)\bigr]\frac{\A\rho^2e^{\im\vartheta}}{2|z|},
\end{equation}
as $|z|\to\infty$, for $\vartheta=\arg z$.  This far-field flow behavior is critically dependent on the effective radius $\rho(w)$, and hence the anchoring strength $w$. The plume angle, thus, decreases from $\pi/4$ to $0$  as the anchoring strength $w$ is increased from $0$ to $\infty$, while the plume speed is largest at $w=8/3$. The velocity field does not appear as a dipole or stresslet in the far-field, but the  sum of a dipole and a quadrupole  --- the mere presence of the body initiates additional flow far away due to the distributed active stress, leading to bulk flow release. This also demonstrates that a body with non-trivial anchoring conditions is sufficient to induce a large scale flow through its presence alone.

\section{Surface tractions and soft boundaries}

Stresses due to passive LC elasticity, actively-driven viscous flow, and direct activity all contribute to surface forcing. The LC introduces a surface traction  $\bm{t}^e= \bm{\n}\cdot\bm{\hat\sigma}^e + \partial_s\bm{t}^s$, where $\bm{\hat\sigma}^e$ is the Ericksen elastic stress tensor given in \eqref{eq:tracelessstress} and 
\begin{equation}
    \bm{t}^s= \frac{w}{2}\sin^2(\theta-\theta_0)\bm{\s}+\frac{w}{2}\sin[2(\theta-\theta_0)]\bm{\n},
\end{equation} 
is a surface stress vector associated with finite surface-anchoring \cite{cs24a}. Activity, meanwhile, produces a surface traction $\bm{t}^a = \Er\,\bm{\n}\cdot(\bm{\hat\sigma}^v+\A\bm{\hat\sigma}^a-p\I)$ for the (traceless) viscous and active stress tensors, $\bm{\hat\sigma}^v$ and $\bm{\hat\sigma}^a$ given in \eqref{eq:tracelessstress}.

If a boundary is soft, it will generally be deformed by the surface tractions above. As a first step towards understanding more complex materials, we assume the boundary is part of an isotropic solid that  deforms according to plane stress/strain in the framework of linear elasticity, with  shear modulus $\mu_s$ and Poisson's ratio $\nu$ \cite{england2003,Howell2009}. At equilibrium, the elastic solid can be described by an Airy-stress function, $\mathcal{A}(x,y)$, which satisfies the biharmonic equation, $\nabla^4\mathcal{A}=0$. We can, thus, write
\begin{equation}\label{eq:GoursatA}
    \mathcal{A} = \Im[\bar{z}G(z)+H(z)],
\end{equation}
for two (new) Goursat functions, $G$ and $H$,  that  are locally holomorphic within the solid (but not necessarily the LC), analogous to $g$ and $h$ in \eqref{eq:psi} \cite{england2003}. These Goursat functions yield the symmetric stress tensor within the solid, $\Sigma_{ij}$,   via the Kolosov--Muskhelishvili formulae,
 \begin{subequations}\label{eq:Airystresses}
 \begin{gather}
 \Sigma_{11}-\Sigma_{22}-2\im\Sigma_{12}= 2\im\bar{z}G''(z)+2\im H''(z)\\
 \text{and} \quad\Sigma_{11} +\Sigma_{22}=4\Im G'(z),
 \end{gather}
 \end{subequations}
and  the solid displacement, $(U,V)$, can be expressed as
\begin{equation}\label{eq:displacements}
    2M(U-\im V) = \im\kappa\overline{G(z)}+\im\bar{z}G'(z)+\im H'(z),
\end{equation}
up to an arbitrary body displacement and rotation, for the dimensionless elastic modulus  $\M\coloneqq\mu_s L^2/K$  and  $\kappa =(3-4\nu)$ or $(3-\nu)/(1+\nu)$, when assuming plane strain or plane stress, respectively.

Stress balance at the LC-solid interface, $\bm{\n}\cdot\bm\Sigma+\bm{t}^e+\bm{t}^a=\bm{0}$, yields the boundary conditions for $G$ and $H$:
\begin{equation}\label{eq:app_stressbalance}
    \partial_s\left[\bar{z}G'(z)-\overline{G(z)}+H'(z)\right]=(t^e_x+t^a_x)-\im (t^a_y+t^e_y).
\end{equation}

\subsection{Activity-controlled deformations}

Returning to the example of a unit cylinder with tangential anchoring, the  director angle \eqref{eq:cylinder_theta} and  streamfunction  \eqref{eq:cylinder_psi} yield analytic expressions for the elastic traction, $t^e_x+\im t_y^e$,  and  activity-induced traction, $t^a_x+\im t_y^a= \A\Er(1/z-\rho^2z)$. The cylinder is now assumed to be deformable, and so, we seek the  Airy stress \eqref{eq:GoursatA}. Here, the Goursat functions $G(z)$ and $H(z)$ must be locally holomorphic in $|z|<1$ and satisfy \eqref{eq:app_stressbalance}  on $|z|=1$; the solution is provided in \cite{SuppMat}.

\begin{figure}[t!]
    \centering
    \includegraphics[width=\linewidth]{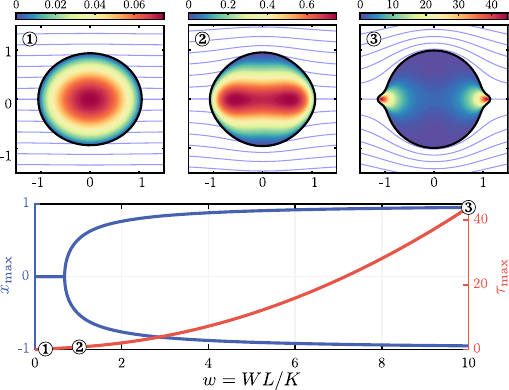}
    \caption{(top) Deformation of a soft cylinder immersed in a passive nematic LC with $\A\Er=0$, $\M/\w=5$, $\kappa=5/3$, and dimensionless anchoring strengths \ding{172}~$\w=0.1$, \ding{173}~$\w=1$, and \ding{174}~$\w=10$. Color  indicates the maximum  shear stress, $\tau=\sqrt{(\Sigma_{11}-\Sigma_{22})^2/4+\Sigma_{12}^2}$, within the cylinder and the blue curves are director lines. (bottom) The positions, $(x_{\max},0)$, and value, $\tau_{\max}=\max_{\bm{x}}\tau$, of the largest maximum shear stress plotted as a function of $w$.}
    \label{fig:Fig2}
\end{figure}

Without activity ($\A=0$), boundary deformations help to  reduce the elastic energy stored in the bulk LC \cite{cs24a}. Figure~\ref{fig:Fig2} shows the LC configurations and body deformations for three different anchoring strengths and $\kappa=5/3$ (\ie~an incompressible cylinder, $\nu=1/2$, under plane stress). For large anchoring strengths, $w\gg1$, most of the LC  energy is stored local to the left and right poles of the cylinder, with a singular energy emerging in the case of strong (infinite) anchoring ($w=\infty$) due to the appearance of two $-1$ Boojum topological defects at $z=\pm1$ \cite{vl83,cs23}. This localized  energy within the LC results in large localized surface tractions  as $w\to\infty$: $t^e_x+\im t_y^e =\Oh(w^2)$ when $z\pm 1=\Oh(1/w)$, and $t^e_x+\im t_y^e=\Oh(1)$ otherwise. This, in turn, leads to large localized deformations; in particular, we find that 
\begin{equation}\label{eq:strongdeformation}
U+\im V\sim  \frac{3w}{8M}\bigl[\mathcal{U}(1-z)-\mathcal{U}(1+z)\bigr],
\end{equation}
 as $w\to\infty$, where 
\begin{equation}\label{eq:strongdeformation_local}
\mathcal{U}(Z/w) =\frac{\kappa}{1+Z}+\frac{1}{1+\bar{Z}}+\frac{Z+\bar{Z}}{(1+\bar{Z})^2},
\end{equation}
corresponds to a deformation that is only apparent when $z\pm 1=\Oh(1/w)$. As the anchoring strength decreases, stress is no longer concentrated at the poles and the maximum shear stress $\tau(x,y)=\sqrt{(\Sigma_{11}-\Sigma_{22})^2/4+\Sigma_{12}^2}$  (a common indicator of mechanical failure \cite{Howell2009}) is greatest at $(\pm x_{\max},0)$, where $x_{\max}$ decreases from $1$ when $w=\infty$, to $0$ when $w\leq 24/35$, as shown in Fig.~\ref{fig:Fig2}.

\begin{figure}[t!]
    \centering
    \includegraphics[width=\linewidth]{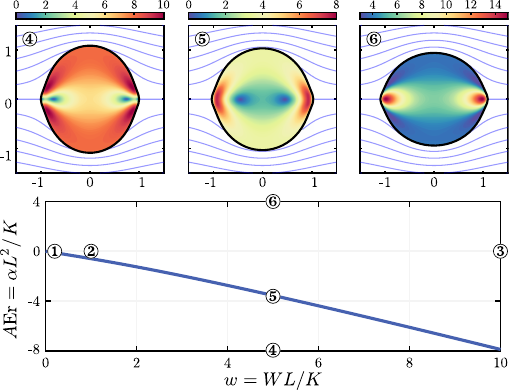}
    \caption{(top) Deformation of a soft cylinder immersed in an active nematic LC with $\w=5$, $\M=50$, $\kappa=5/3$, and \ding{175}~$\A\Er=-8<\mathcal{W}$ (extensile), \ding{176}~$\A\Er=\mathcal{W}\approx -3.55$ (extensile), and \ding{177}~$\A\Er=4>\mathcal{W}$ (contractile). Color indicates the maximal  shear stress, $\sqrt{(\Sigma_{11}-\Sigma_{22})^2/4+\Sigma_{12}^2}$, within the cylinder and the blue curves are director lines. (bottom) The critical active Ericksen number, $\A\Er = \mathcal{W}$, at which the deformed cylinder has aspect ratio one, plotted as a function of $\w$  for $\kappa=5/3$. The points  \ding{172}--\ding{174} correspond to the configurations shown in Fig.~\ref{fig:Fig2}.}
    \label{fig:Fig3}
\end{figure}

In an active nematic ($A\neq 0$), the activity leads to an additional deformation: 
\begin{equation}
    U+\im V= U^e+\im V^e +  \frac{\A\Er}{4\M}\bigl[(1-\kappa)\rho^2z+2\bar{z}\bigr],
\end{equation}
where $(U^e, V^e)$ is the deformation of a cylinder immersed in a passive LC. As this addition is proportional to the activity strength, $\A$, activity can  act with or against the elastic effects of the LC, depending on if it is extensile ($\A<0$) or contractile ($\A>0$). In particular, the aspect ratio, $\mathcal{A}_R$, of the deformed cylinder is given by
\begin{equation}
 \mathcal{A}_R\sim 1+ U\bigr|_{z=1}-V\bigr|_{z=\im} = 1+\frac{\A\Er-\mathcal{W}}{\M},
\end{equation}
where 
\begin{equation}
    \mathcal{W}=\frac{3\kappa}{2\rho}\left(\arctanh \rho-\arctan\rho\right)-\frac{3w}{8}(1+\kappa).
\end{equation}
Thus, if  $\A\Er>\mathcal{W}$ (which includes passive LCs since $\mathcal{W}<0$), the deformed shape is elongated with the  $x$-axis (\ie~the preferred axis of the LC). While, if  $\A\Er<\mathcal{W}$, the shape is elongated with the $y$-axis (\ie~perpendicular to the preferred axis). The aspect ratio is one at the critical value $\A\Er=\mathcal{W}$,  see Fig.~\ref{fig:Fig3}. Therefore, contractile activity ($\A>0$), in this setting, always leads to parallel elongation, while extensile activity ($\A<0$) yields perpendicular elongation, as long as the activity is sufficiently strong. 

\subsection{Bounded domains and spontaneous stirring}
 
 The fluid flow triggered by the presence of a boundary may also be advantageous in interior domains like biological cells --- settings in which a distribution of molecules, such as nutrients, chemicals, or oxygen, benefit the system's function \cite{gtv08,fwwpkms14,kct17,mbk20}. Consider an active nematic LC bounded inside and (weakly) tangentially anchored to a cylinder with anchoring strength $w$. Here, the director angle of the LC can be written as $\theta=\arg(1-\rho^2z^2)$, for the effective radius introduced in the external problem, $\rho(w) =({\sqrt{1+4/\w^2}-2/\w})^{1/2}$, while   the   flow can  be expressed in terms of a streamfunction (see \cite{SuppMat}):
\begin{equation}
\begin{split}
    \psi = \frac{\A}{8}(|z|^2-1)\Im\Big[\frac{1-\rho z}{z^2}&\log(1-\rho z)\\
    +&\frac{1+\rho z}{z^2}\log(1+\rho z)\Big].
\end{split}
\end{equation}

\begin{figure}[ht!]
    \centering
    \includegraphics[width=\linewidth]{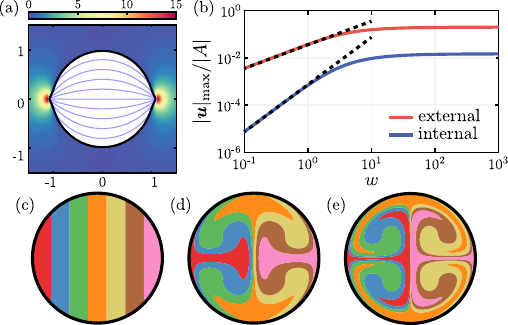}
    \caption{(a) Deformation of a soft cylinder enclosing a passive nematic LC with $\A\Er=0$, $w=10$, $M=50$, and $\kappa=5/3$. Color  delineates the maximum  shear stress, $\sqrt{(\Sigma_{11}-\Sigma_{22})^2/4+\Sigma_{12}^2}$, within the solid and the blue curves are director lines.  (b) Plot of the maximum fluid speed, $|\bm{u}|_{\max}=\max_{\bm{x}}\sqrt{u^2+v^2}$, as a function of $\w$ for the  LC external  and internal to a cylinder. The asymptotic behaviours  as $w\to 0$  are shown as  black dashed lines.  (c--e) Snapshots at times  (c) $t=0$,  (d) $t=100/|A|$, and (e) $t=200/|A|$  of colored fluid particles transported  due to extensile activity inside the cylinder for $w=10$.}
    \label{fig:Fig4}
\end{figure}

Figure~\myref{fig:Fig4}{a} shows the director lines inside a  cylinder that has deformed according to linear elasticity \cite{SuppMat}. The final deformed shape resembles a tactoid \cite{sv22}, with a localized deformation at the left and right poles occurring for large anchoring strengths due to the appearance of two $+1$ aster defects when $w=\infty$. The deformed shape is  dependent on the active Ericksen number, $\A\Er$, with extensile activity ($A<0$) leading to parallel elongation  and  sufficient contractile activity ($A<0$) yielding perpendicular elongation \cite{SuppMat} --- opposite to the  dependence on the activity strength  in the external problem. 

Figures~\myref{fig:Fig4}{c--e} show stirring induced by extensile activity. Varying the activity and anchoring strength affects the rate of stirring. The maximum fluid speed, $|\bm{u}|_{\max}$, is plotted as function of anchoring strength in Fig.~\myref{fig:Fig4}{b}, for both the external and internal problems. The speeds plateau for large anchoring strengths, and decrease according to  $|\bm{u}|_{\max}\sim 2 |\A|\w/(25\sqrt{5}) $ and $|\bm{u}|_{\max}\sim  |\A|\w^2/(600\sqrt{5})$ as $w\to 0$, for the external  and internal problem, respectively. The extent to which this stirring flow is mixing is an interesting but separate question \cite{Villermaux19}. Mixing is notoriously challenging in highly viscous flows and in confined geometries \cite{gkddrt07}, but is likely promoted by a wide spatial distribution of active stress \cite{trsoafmh19,mhsemh24}. 

\section{Conclusions and discussions}

We have shown that the configurations and dynamics of an active nematic can be affected dramatically by the mere presence of a boundary, for any non-trivial anchoring strength. It suggests that energy bound in a uniform active nematic can be released as a large scale flow due to a localized inclusion alone.

The spontaneous flow that arises is characterized by unique plumes whose structure is anchoring strength dependent. The body may then deform in response to this anisotropic, viscoelastic flow, in a generally localized manner near (virtual) topological defects. Such flows can also produce a spontaneous stirring, and likely mixing, of the environment. The associated symmetry breaking can lead to cell or droplet division and motility \cite{tmc12,gd14,kfpsv22}. Looking to the horizon, biological cells are host to a multitude of active anisotropic networks. Can such fluids be triggered into motion solely by the passive anchoring conditions on an arriving organelle? This mechanism could offer an energy-efficient way to generate precisely timed mixing flows within the cellular environment.

New experiments on the shapes and dynamics of deformable inclusions in active nematics is needed. Testable hypotheses include the differential direction of elongation, which may be tuned by varying either the activity strength (\eg~by changing the density of molecular motors ~\cite{ojb20}) or by changing the body stiffness. With sufficient perpendicular elongation due to strong extensile activity, as in the top left panel of Fig.~\ref{fig:Fig3}, body rotation may also ensue to reduce the bulk elastic energy. We thus predict that a diverse zoology of body dynamics will emerge with increasing extensile activity in particular.

Consequences for intracellular form and function are also expected. Taking the metaphase spindle as an example \cite{dmmw98,onb18}, using a length scale $L=\SI{20}{\micro m}$, bulk elastic constant $K=\SI{400}{p N}$ \cite{ojb20}, and contractile activity strength $\alpha \sim\SI{35}{pN/\micro m^2}$ 
\cite{smilk11,bn14,ojb20}, we find that $A\mathrm{Er}\approx 35$. Balancing against surface tension, comparable parameters can provide the spindle's characteristic shape \cite{bn14,ojb20}. Such elastic and activity-induced stresses would both stretch an immersed body in the direction parallel to the background director field. These mechanics would seem to be of additional service in the splitting of sister chromatids, which separate and are pulled towards the two spindle poles during anaphase A \cite{rrsewvss04,yrwkycfmn19}. Chromosomes can also be repelled in a direction perpendicular to the director in metaphase, due to LC elastic interactions \cite{cs24,krn24}. Our results suggest a more complex picture if active stresses are included, which would impose an additional attraction (\eg~from Fig.~\ref{fig:Fig1} with oppositely signed flow).

Extensile activity in a related synthetic system, taking $L\approx \SI{200}{\micro m}$, instead has $A\mathrm{Er}\approx -1500$ \cite{rzd23,scdhd12,Thampi2013}. Such a highly extensile suspension is likely to result in large deformations of any soft inclusions in the direction perpendicular to the background director field. This could add to the already complex body dynamics that have been observed in such systems \cite{rzd23}.





Future directions will require different machinery.  Large deformations will more completely couple the LC configuration and flows, and a finite rotational viscosity will introduce an additional timescale \cite{degennes1993,Rey10}. The analytical limits provided herein may still be of use for steering investigations of these more highly nonlinear regimes.






\begin{acknowledgments}
We acknowledge helpful discussions with Nicholas L.\ Abbott and Ido Lavi. Support for this research was provided by the Office of the Vice Chancellor for Research and Graduate Education with funding from the Wisconsin Alumni Research Foundation.
\end{acknowledgments}

\bibliography{apssamp}
\end{document}


\preprint{APS/123-QED}

\title{Supplemental material for ``Active nematic response to a deformable body or boundary: elastic deformations and anchoring-induced flow''}

\author{Thomas G.~J.~Chandler}
\email{tgchandler@wisc.edu}
\author{Saverio E.~Spagnolie}%
\email{spagnolie@wisc.edu}
\affiliation{%
Department of Mathematics, University of Wisconsin–Madison, Madison, WI 53706, USA}%

\date{\today}

\maketitle

\section{Governing equations}

According to Erickson--Leslie theory \cite{stewart2004}, in the absence of any body forces and torques, the components (denoted by subscripts) of the director field, $n_i$, and fluid velocity, $u_i$, evolve according to 
\begin{subequations}\label{eq:EL}
\begin{gather}
   \gamma_1 N_i=H_i -\gamma_2 E_{ik}n_k-L n_i,\\
    \rho(\partial_tv_i+v_k\partial_{k}v_i) = \partial_k\sigma_{ki},\quad\text{and}\quad   \partial_iu_i=0,\subtag{b,c}
\end{gather}
\end{subequations}
for the co-rotational time flux $N_i=\partial_tn_i+v_k\partial_{k}n_i+\Omega_{ik}n_k$; Lagrange multiplier $L$, which imposes $n_in_i=1$;  strain-rate tensor and vorticity tensor
\begin{subequations}
\begin{equation}
    E_{ij}=\frac{1}{2}(\partial_iu_j+\partial_j u_i)\quad\text{and}\quad  \Omega_{ij}=\frac{1}{2}(\partial_iu_j-\partial_j u_i),\subtag{a,b}
\end{equation}
\end{subequations}
respectively;
Ericksen torque stress tensor and molecular field
 \begin{subequations}
\begin{equation}
    \Pi_{ij}=\frac{\partial \mathcal{F}}{\partial\partial_i n_j}\quad\text{and}\quad H_i =\partial_k\Pi_{ki} -\frac{\partial \mathcal{F}}{ \partial n_i},\subtag{a,b}
\end{equation}
\end{subequations}
respectively, where $\mathcal{F}(\bm{n},\nabla\bm{n})$ is the Frank free energy of the LC; and  stress tensors 
\begin{subequations}\label{eq:stressEL}
\begin{align}
   \sigma_{ij} &=(\mathcal{F}-p)\delta_{ij} -\partial_i n_k \Pi_{jk} +2\alpha n_in_j+\tilde{\sigma}_{ij},\\
   \tilde{\sigma}_{ij}&=\alpha_1n_kE_{kp}n_p n_in_j+\alpha_2 N_in_j+\alpha_3 n_i N_j+\alpha_4 E_{ij} +\alpha_5 n_jE_{ik}n_k+\alpha_6 n_i E_{jk}n_k,
\end{align}
\end{subequations}
where repeated indices denote summation and $\delta_{ij}$ is the Kronecker delta. Note that $\sigma_{ij}$ is not a symmetric tensor, and it is defined such that $\nu_i\sigma_{ij}$ is the $j$th component of the stress acting on a surface of normal $\bm{\hat\nu}$.   Here,  $\rho$ is the density, $p$ is the pressure,  $\alpha$ is the activity strength, $\alpha_i$ are the Leslie viscosities,   $\gamma_1=\alpha_3-\alpha_2$, and $\gamma_2= \alpha_3+\alpha_2=\alpha_6-\alpha_5$, where the last equality is the Parodi relation. Under the one-constant approximation, $\mathcal{F}=(K/2)\lVert\nabla\bm{n}\rVert^2\equiv(K/2)\partial_in_j\partial_in_j$ for the single (Frank) elastic constant $K$, this yields the torque stress tensor $\Pi_{ij}=K\partial_i n_j$ and molecular field $H_{i}=K\nabla^2 n_i$.

Using the fact that $N_in_i=0$ (which follows from $n_in_i=1$), the evolution equation \myeqref{eq:EL}{a} can be decomposed into its component parallel and perpendicular to $\bm{n}$,
\begin{subequations}\label{eq:EL2}
\begin{equation}
     L=-\gamma_2 n_kE_{ki}n_i\quad\text{and}\quad \gamma_1 N_i = h_i-\gamma_2n_kE_{kj}(\delta_{ji}-n_jn_i),\subtag{a,b}
\end{equation}
\end{subequations}
respectively, where $h_i=(\delta_{ij}-n_in_j)H_j$ is the projected molecular field. Inserting \myeqref{eq:EL2}{b} into \myeqref{eq:stressEL}{b} and rearranging yields
\begin{subequations}
\begin{equation}
   \tilde{\sigma}_{ij}=\alpha_4 E_{ij} +\left(\alpha_1+\frac{\gamma_2^2}{\gamma_1}\right)n_kE_{kp}n_p n_in_j+\frac{\alpha_5\alpha_3-\alpha_6\alpha_2}{\gamma_1}\left( n_i E_{jk}n_k+n_jE_{ik}n_k\right)+\frac{\alpha_2}{\gamma_1} h_jn_j+\frac{\alpha_3}{\gamma_1} n_i h_j.
\end{equation}
\end{subequations}
Consequently, we introduce five new  constants (to replace the five independent viscosities above),
\begin{subequations}
\begin{equation}
    \mu\coloneqq \frac{\alpha_4}{2},\quad \mu_1\coloneqq \alpha_1+\frac{\gamma_2^2}{\gamma_1},\quad \mu_2\coloneqq\frac{\alpha_5\alpha_3-\alpha_6\alpha_2}{\gamma_1},\quad\lambda\coloneqq -\frac{\gamma_2}{\gamma_1},\quad \gamma\coloneqq \gamma_1. \subtag{a--e}
\end{equation}
\end{subequations}
Assuming the system's Reynolds number is negligible (\ie~the left-hand side of  \myeqref{eq:EL}{b} is zero), we are then left with the system introduced in the main text, Eqs.~(1)--(3).

Non-dimensionalizing this system using a length scale $L$, velocity scale $U$, and  time scale $L/U$ yields 
\begin{subequations}\label{eq:dimensionlesseqs}
\begin{gather}
\partial_t\bm{n}+\bm{u}\cdot\nabla\bm{n} =\bm{n}\cdot\W+ \lambda\bm{n}\cdot \E\cdot(\I-\bm{n}\bm{n})+\frac{\bm{h}}{\gamma'\Er},\\
   \nabla\cdot\left(\bm\sigma^e+\Er\, \bm\sigma^v+\A\Er\,\bm\sigma^a- \Er\, p \I\right)=\bm{0}\quad\text{and}\quad \nabla\cdot\bm{u}=0,\subtag{b,c}
\end{gather}
\end{subequations}
for the elastic, viscous, and active stress tensors
\begin{subequations}
\begin{align}
   \bm \sigma^e &=\frac{1}{2}\lVert\nabla\bm{n}\rVert^2\I -\nabla \bm{n}\cdot \nabla \bm{n}^T -\frac{1}{2}\left(\bm{n}\bm{h}-\bm{h}\bm{n}\right) -\frac{\lambda}{2}\left(\bm{n}\bm{h}+\bm{h}\bm{n}\right),\\
    \bm \sigma^v &=2\E+\mu_1'(\bm{n}\cdot\E\cdot\bm{n}) \bm{n}\bm{n}+\mu_2'(\bm{n}\E\cdot\bm{n}+\bm{n}\cdot\E\bm{n}),\\
    \bm  \sigma^a &= 2 \bm{n}\bm{n},
\end{align}
\end{subequations}
respectively. Here,  $\A$, $\Er$, $\lambda$, $\gamma'$, $\mu_1'$, and $\mu_2'$ are the dimensionless parameters defined in the main text and $\bm{n}\bm{n}$ and similar terms are  dyadic products.

For small viscosity ratios and rotational Erickson number, $\gamma',\mu_i',\gamma'\Er\ll1$, the stress tensors reduce to
\begin{subequations}\label{eq:asymstress}
\begin{align}
   \bm \sigma^e &=\frac{1}{2}\lVert\nabla\bm{n}\rVert^2\I-\nabla \bm{n}\cdot \nabla \bm{n}^T +\Oh(\gamma'\Er),\\
    \bm \sigma^v &=\nabla\bm{u}+\nabla\bm{u}^T+\Oh(\mu_i'\Er),\\
    \bm  \sigma^a &= 2 \bm{n}\bm{n}.
\end{align}
\end{subequations}
Inserting these into \eqref{eq:dimensionlesseqs} yields the asymptotic system  found in the main text
\begin{subequations}\label{eq:system_smallEr}
\begin{gather}
\bm{h} = (\I-\bm{n}\bm{n})\cdot\nabla^2\bm{n} = \Oh(\gamma'\Er),\quad \nabla\cdot\bm{u}=0,\subtag{a,b}\\
  \nabla^2\bm{u}-\nabla p+2A\nabla\cdot(\bm{n}\bm{n})=  \Oh(\gamma',\mu_i').\subtag{c}
\end{gather}
\end{subequations}
where we have used that $\nabla\cdot\bm{\sigma}^e=\Oh(\gamma'\Er)$, which follows from \myeqref{eq:system_smallEr}{a} and $|\bm{n}|=1$.

\subsection{Two-dimensional theory}

In 2D,  the LC can be described by a director angle,  $\theta(x,y)$, and a  streamfunction, $\psi(x,y)$, such that $\bm{n} = (\cos\theta,\sin\theta)$ and $\bm{u}=(\psi_y,-\psi_x)$.  Here, we  absorb the isotropic parts of the stress tensors, \eqref{eq:asymstress}, into the pressure by taking $p\mapsto p+\A$.  At leading-order, the  system \eqref{eq:system_smallEr} is 
\begin{subequations}\label{eq:asym_sym2d}
\begin{gather}
\qquad\nabla^2\theta=0,\\
\partial_y\omega-\partial_xp=\A \left[\partial_x\cos(2\theta)+\partial_y\sin(2\theta)\right],\\
\partial_x\omega+\partial_y p=\A \left[\partial_y\cos(2\theta)-\partial_x\sin(2\theta)\right],
\end{gather}
\end{subequations}
with vorticity $\omega = -\nabla^2\psi$. On any immersed or confining boundaries, the fluid velocity vanishes and the director field is subject to Rapini–Papoular  anchoring~\cite{rp69,cs23}, \ie
\begin{subequations}\label{eq:2DBCs}
\begin{equation}
\partial_s\psi=\partial_\nu\psi= 0\quad \text{and}\quad \partial_{\nu}\theta=\frac{w}{2}\sin\left[2(\theta-\theta_0)\right], \subtag{a,b}
\end{equation}
\end{subequations}
respectively, where $w$ is the dimensionless anchoring strength (as defined in main text), $\theta_0$ is the preferred director orientation on the boundary, and $\partial_s=\bm{\s}\cdot\nabla$ and $\partial_\nu=\bm{\n}\cdot\nabla$ are the arclength derivative and normal  derivative, respectively, for the  unit anticlockwise tangent vector $\bm{\s}$ and unit  normal vector pointing into the LC $\bm{\n}$. 

The traceless stress tensors are denoted with hats:
\begin{subequations}\label{eq:asymstress_traceless}
\begin{align}
   \hat{\bm{\sigma}}^e&=\frac{1}{2}[(\partial_y\theta)^2-(\partial_x\theta)^2](\bm{x}\bm{x}-\bm{y}\bm{y}) -\partial_{x}\theta\partial_y\theta (\bm{x}\bm{y}+\bm{y}\bm{x}),\\
    \hat{\bm{\sigma}}^v &= 2\partial_{xy}\psi(\bm{x}\bm{x}-\bm{y}\bm{y})+(\partial_{yy}\psi-\partial_{xx}\psi)(\bm{x}\bm{y}+\bm{y}\bm{x}),\\ 
    \hat{\bm{\sigma}}^a&= \cos(2\theta)(\bm{x}\bm{x}-\bm{y}\bm{y})+\sin(2\theta)(\bm{x}\bm{y}+\bm{y}\bm{x}),
\end{align}
\end{subequations}
where $\bm{x}$ and $\bm{y}$ are the unit vectors in the $x$ and $y$ directions, respectively. The elastic and activity-induced stresses  acting on a surface  are  
\begin{subequations}
\begin{equation}
\bm t^e =\bm{\n}\cdot\hat{\bm{\sigma}}^e+\partial_s \bm{t}^s\quad\text{and}\quad
\bm t^a=\Er\,\bm{\n}\cdot(\hat{\bm{\sigma}}^v+\A\hat{\bm{\sigma}}^a-p\I),
\end{equation}
\end{subequations}
 respectively, for the surface stress vector (which comes from the surface anchoring condition \cite{cs24a})
\begin{equation}
    \bm{t}^s=\frac{w}{2}\sin(\theta-\theta_0)^2 \bm{\s}+\frac{w}{2}\sin\left[2(\theta-\theta_0)\right]\bm{\n} .
\end{equation}

Changing coordinates from $x$ and $y$ to the complex position $z=x+\im y$ and its complex conjugate $\bar{z}=x-\im y$ yields the problem
\begin{subequations}
\begin{equation}
\partial_{z\bar{z}}\theta=0\quad\text{and}\quad
\partial_{\bar{z}}(p-\im\omega)=\A \partial_{z}e^{2\im\theta}.\subtag{a,b}
\end{equation}
\end{subequations}
Integrating yields the solution
\begin{subequations}\label{eq:tpompsi}
\begin{equation}
    \theta=-\arg f'(z), \quad p-\im\omega =\A \partial_z\bigl[g(z)+ \overline{f(z)}/f'(z)\bigr], \quad \psi =\frac{\A}{4}\Im\Bigl[\bar{z}g(z)+h(z)+\overline{\int f(z)\de z}\,/f'(z)\Bigr],\subtag{a--c}
\end{equation}
\end{subequations}
for  functions $f$, $g$, and $h$, which are locally holomorphic in the LC and satisfy the boundary conditions \eqref{eq:2DBCs}. Note that the fluid velocity, $\bm{u}=(u,v)$, can be expressed as
\begin{equation}
    u-\im v = 2\im \psi_z = \frac{\A }{4}\left[\bar{z}g'(z)-\overline{g(z)}+h'(z)+\partial_z\left(\frac{\overline{\int f(z)\de z}}{f'(z)}\right)-\frac{f(z)}{\overline{f'(z)}}\right].
\end{equation}
On a bounding surface, $\bar{z}$ can be related to $z$ using the boundary's Schwarz function \cite{ablowitz2003}, and the surface tractions can  be computed using
\begin{subequations}\label{eq:app_trac_complex}
\begin{align}
t^e_x-\im t^e_y&=  2\im(\partial_z\theta)^2 \partial_s z+\partial_s(t^s_x-\im t^s_y),\\
t^a_x-\im t^a_y&= \Er \bigl([4\partial_{zz}\psi  -\im \A e^{-2\im\theta}]\partial_sz- \im p\,\partial_s\bar{z}\bigr),\\
t^s_x-\im t^s_y&=\frac{w}{8}\bigl(2+\e^{2\im(\theta-\theta_0)}-3\e^{2\im(\theta_0-\theta)}\bigr)\partial_s \bar{z}.
\end{align}
\end{subequations}

\subsection{Solid deformation}

According to plane stress/strain in linear elastostatics,  the deformation, $(U,V)$, and stress tensor, $\bm\Sigma$, of an isotropic elastic solid can  be written in terms of an Airy stress function, $\mathcal{A}(x,y)$, which satisfies the biharmonic equation, $\nabla^4\mathcal{A}=0$ \cite{england2003,Howell2009}. This biharmonic function can be written in terms of two Goursat functions, $G$ and $H$, that is
\begin{equation}\label{eq:GH}
    \mathcal{A}= \Im\left[\bar{z}G(z)+H(z)\right],
\end{equation}
where $G$ and $H$  are locally holomorphic in the solid and satisfy the stress balance on the LC--solid interface, $-\bm{\n}\cdot\bm\Sigma=\bm{t}^e +\bm{t}^a$. The solid stress tensor, $\Sigma_{ij}$,  is  given by the Kolosov--Muskhelishvili formulae,
\begin{subequations}\label{eq:Airystresses}
\begin{equation}
\Sigma_{11}-\Sigma_{22}-2\im\Sigma_{12}= 2\im\bar{z}G''(z)+2\im H''(z)
 \quad\text{and} \quad\Sigma_{11} +\Sigma_{22}=4\Im G'(z),\subtag{a,b}
\end{equation}
\end{subequations}
and the solid displacement, $(U,V)$, is given by
\begin{equation}\label{eq:displacements2}
    2M(U-\im V) = \im\kappa\overline{G(z)}+\im\bar{z}G'(z)+\im H'(z),
\end{equation}
where $M$ and $\kappa$ are the dimensionless elastic modulus and  Muskhelishvili constant defined in the main text. Using \eqref{eq:app_trac_complex} and \eqref{eq:Airystresses}, the first integral of the interface stress balance  is
\begin{equation}\label{eq:app_stressbalance}
    \bar{z}G'(z)-\overline{G(z)}+H'(z)=\int t^e_x-\im t^e_y+t^a_x-\im t^a_y\de s,
\end{equation}
where $\bar{z}$ can be related to $z$ using the Schwarz function for the interface. This stress balance  is unique up to an arbitrary integration constant.

\section{Example 1:  Cylinder immersed in an active  LC}

In this section, we derive  analytical expressions for the director field, streamfunction, and Airy stress function associated with the  first example considered in the main text: a soft unit cylinder immersed in a weakly-active nematic LC, subject to (finite) tangential anchoring.

\subsection{Director field} 

Chandler and Spagnolie~\cite{cs23} showed that the director field can be expressed in terms of a potential function $f(z)$ such that $\theta=-\arg f'$. Here, $f(z)$ must be  holomorphic in $\rho<|z|<\infty$,  have a constant imaginary-part on $|z|=\rho$, and satisfy $f(z)\sim z$ as $|z|\to\infty$, for the effective/virtual radius $\rho(w)= ({\sqrt{1+4/\w^2}-2/\w})^{1/2}$. By Milne-Thomson circle theorem, this has solution 
\begin{equation}\label{eq:appf}
    f(z)=z+\rho^2/z.
\end{equation}
(Note that this is unique up to an additive imaginary logarithm, $\im\gamma\log z$ for $\gamma\in \mathbb{R}$. We shall only consider the up--down symmetric case with  $\gamma=0$, which minimizes the (Frank) elastic energy \cite{cs23}.) This potential  yields the director angle given in the main text,
\begin{equation}
    \theta=-\arg f'(z) = -\arg(1-\rho^2/z^2),
\end{equation}
which  holds for all anchoring strengths $w> 0$.

\subsection{Streamfunction} 
Inserting \eqref{eq:appf} into \myeqref{eq:tpompsi}{c} yields the streamfunction
\begin{equation}
\psi =\frac{\A }{4}\Im\left[\bar{z}g(z)+h(z)+\frac{\bar{z}^2/2+\rho^2\log\bar{z}}{1-\rho^2/z^2}\right],
\end{equation} 
for the two Goursat functions $h(z)$ and $g(z)$, which are locally holomorphic in $1<|z|<\infty$ and yield a fluid velocity that vanishes on $|z|=1$ and as $|z|\to\infty$. 

Firstly, to ensure $\psi$ is single-valued around $|z|=1$, let $h(z)= \tilde{h}(z)+\rho^2\log z/(1-\rho^2/z^2)$:
\begin{equation}\label{eq:app_psi}
\psi =\frac{\A }{4}\Im\left[\bar{z}g(z)+\tilde{h}(z)+\frac{\bar{z}^2/2+2\rho^2\log|z|}{1-\rho^2/z^2}\right].
\end{equation} 
We now need to find $g$  and $\tilde{h}$ such that the fluid velocity,
\begin{equation}\label{eq:app_uv}
u-\im v=2\im \partial_z\psi=\frac{\A}{4 }\left[\bar{z}g'(z)+\tilde{h}'(z)-\bar{g}(\bar{z})-\frac{z+\rho^2/z}{1-\rho^2/\bar{z}^2}-\frac{2\rho^2}{z^3}\frac{(\bar{z}^2-z^2)/2+\rho^2/2+2\rho^2\log|z|}{(1-\rho^2/z^2)^2}\right],
\end{equation} 
vanishes on $|z|=1$  and as $|z|\to\infty$.  Note that the pressure and vorticity are given by   \myeqref{eq:tpompsi}{b}, which is
\begin{equation}
p-\im \omega=\A \left(g'(z) +\partial_z\frac{\bar{z}+\rho^2/\bar{z}}{ 1-\rho^2/z^2}\right).
\end{equation}

Consider the functions
\begin{subequations}\label{eq:app_goursat}
\begin{equation}
    g(z)= \mathcal{G}(z)-\frac{\rho^2(1+\rho^2 z^2)}{z(z^2-\rho^2)}\quad\text{and}\quad
    \tilde{h}(z)=\mathcal{H}(z)+\frac{z^2}{2}-\frac{1}{2z^2}+\frac{1/2+\rho^4}{z^2-\rho^2}.\subtag{a,b}
\end{equation}
\end{subequations}
These are  holomorphic in $1<|z|<\infty$ provided $\mathcal{G}$ and $\mathcal{H}$ are  holomorphic there. Furthermore,
\begin{subequations}
\begin{align}
  u-\im v&= \frac{\A}{4 }\left[\mathcal{G}'(z)/z+\mathcal{H}'(z)-\bar{\mathcal{G}}(1/z)\right] \enspace \text{on}\enspace |z|=1,\\
\text{and}\quad u-\im v&\sim \frac{\A}{4 }\left[\bar{z}\mathcal{G}'(z)+\mathcal{H}'(z)-\bar{\mathcal{G}}(\bar{z})\right]\enspace\text{as}\enspace|z|\to\infty,
\end{align}
\end{subequations}
which both vanish provided $\mathcal{H}'(z)=\bar{\mathcal{G}}(1/z)-\mathcal{G}'(z)/z$,   $\mathcal{G}(z) \sim a z+b$ as $|z|\to\infty$, and $\mathcal{H}'(z)\sim \bar{b}$ as $|z|\to\infty$, for $a\in\mathbb{R}$ and $b\in \mathbb{C}$. Since $\mathcal{G}$ and $\mathcal{H}$ are holomorphic in $1<|z|<\infty$, they can be expressed as Laurent expansions around $z=0$. It immediately follows that  $\mathcal{G}(z)= a z+b$ and $\mathcal{H}(z) =  \bar{b}z+c$ for $c\in\mathbb{C}$. Here, $b$ and $c$ are  complex constants that can be set to zero, without loss of generality,  while   $a$ is a real constant that sets the  reference pressure at infinity,
\begin{equation}
p-\im\omega \sim \A a \enspace\text{as}\enspace |z|\to\infty,
\end{equation} 
which we also take to be zero.

Inserting \eqref{eq:app_goursat} into \eqref{eq:app_psi} and \eqref{eq:app_uv} with $\mathcal{G}=\mathcal{H}=0$  yields the streamfunction  given in the main text,
\begin{equation}
    \psi= \frac{\A\rho^2}{8}\Im\left[\frac{(|z|^2-1)^2}{z^2(z^2-\rho^2)}+\frac{1-|z|^2+2\log |z|}{(z^2-\rho^2)/(2\rho^2)}\right],
\end{equation}
and the fluid velocity
\begin{equation}
u-\im v=2\im \partial_z\psi=\frac{\A\rho^2}{4 }(|z|^2-1)\left[\frac{2/z-\bar{z}+\rho^2 (z-1/z^3)+\rho^4/z}{(z^2-\rho^2)^2}-\frac{1/\bar{z}-\rho^2/z}{\bar{z}-\rho^2}\right]-\frac{\A \rho^4 z\log|z|}{(z^2-\rho^2)^2}.
\end{equation} 
Furthermore, using $z_s=\im z$, $\bar{z}=1/z$, and  $w =4\rho^2/(1-\rho^4)$, the  tractions \eqref{eq:app_trac_complex} acting on $|z|=1$ are
\begin{subequations}\label{eq:cylinderin_tractions}
\begin{equation}
    t_x^e-\im t_y^e = \frac{3z(1/\rho^2+z^2)}{2(1/\rho^2-z^2)^2}- \frac{3z(z^2-3\rho^2)}{2(z^2-\rho^2)^2}+\frac{3+\rho^2}{1+\rho^2}\frac{1}{2z} \quad\text{and} \quad t_x^a-\im t_y^a = \A \Er\Big(z-\frac{\rho^2}{z}\Big). \subtag{a,b}
\end{equation}
\end{subequations}

\subsection{Airy stress function} 

We next find the solid Goursat functions $G(z)$ and $H(z)$, which must be holomorphic in $|z|<1$ and satisfy the stress balance  \eqref{eq:app_stressbalance} on $|z|=1$. Integrating the surface tractions given in \eqref{eq:cylinderin_tractions}, and imposing  $\bar{z}=1/z$ and $z_s=\im z$, we find that the stress balance can be expressed as
\begin{equation}\label{eq:cylinderin_stressbalance}
\frac{1}{z}G'(z)-\overline{G}(1/z)+H'(z)=    \frac{\im (3+\rho^2)}{2z(1+\rho^2)}+\frac{3\im z/2}{z^2-\rho^2}+\frac{3\im z/2}{z^2-1/\rho^2}+\frac{3\im}{2\rho}\log\frac{1-\rho/z}{1+\rho/z} -\im A\Er\Big(z+\frac{\rho^2}{z}\Big).
\end{equation}
The trick to finding $G$ and $H$ here is to use that $G(z)$ and $H'(z)$ are holomorphic in $|z|<1$, while $\bar{G}(1/z)$ is holomorphic in $|z|>1$; this allows one to match the terms on the left-hand side of \eqref{eq:cylinderin_stressbalance} with those on the right-hand side.

Consider the functions
\begin{subequations}\label{eq:app_Goursat_Airy}
\begin{align}
    G(z) &=\mathcal{G}(z)+\frac{3+2\rho^2}{1+\rho^2}\frac{\im z}{2}+\frac{3\im z/(2\rho^2)}{1/\rho^2-z^2}-\frac{3\im}{2\rho}\log\frac{1/\rho +z}{1/\rho -z}-\im\A\Er \frac{\rho^2 z}{2},\\
 \text{and} \quad  H(z) &=\mathcal{H}(z)-\frac{3\im z^2/2}{1/\rho^2-z^2}-\im\A\Er \frac{z^2}{2},
\end{align}
\end{subequations}
These are holomorphic in $|z|<1$ as long as $\mathcal{G}$ and $\mathcal{H}$ are holomorphic there, and \eqref{eq:cylinderin_stressbalance} is satisfied provided $\mathcal{H}'(z)=\bar{\mathcal{G}}(1/z)-\mathcal{G}'(z)/z$.  As $\mathcal{G}$ and $\mathcal{H}$ are holomorphic at $z=0$, they can be expressed as Taylor expansions around $z=0$. It follows that $\mathcal{G}(z)=a z+b$ and $\mathcal{H}(z)=\bar{b}z+c$ for $a\in\mathbb{R}$ and $b,c\in \mathbb{C}$. Constant  $c$ is arbitrary and can be set to zero,  while $b$ and $a$ correspond to a rigid body motion. In particular, using the displacement equation \eqref{eq:displacements2} with \eqref{eq:app_Goursat_Airy}, the origin is rotated anticlockwise by the signed angle $-a(1+\kappa)/(2M)$ and is displaced by $U_0+\im V_0=\im\bar{b}(1+\kappa)/(2 M)$. We  set  $a$ and $b$  to zero so that the origin and its orientation are preserved. 

Overall, inserting \eqref{eq:app_Goursat_Airy} with $\mathcal{G}=\mathcal{H}=0$ into \eqref{eq:Airystresses} and \eqref{eq:displacements2} gives the stress and  displacement of the cylinder, respectively. The deformed cylinder's aspect ratio and  the asymptotic displacement as $w\to\infty$ given in the main text  follow from these expressions.

\section{Example 2: Cylinder confining an active LC}

In this section, we derive  analytical expressions for the director field, streamfunction, and Airy stress function associated with the  second example considered in the main text: a weakly-active nematic LC inside a soft unit cylinder, subject to (finite) tangential anchoring.

\subsection{Director field}
Using the effective/virtual boundary technique presented in Ref.~\cite{cs23},  the director angle can be described by a potential function $f(z)$, which is holomorphic in $|z|<1/\rho$ and  has a constant imaginary-part on $|z|=1/\rho$,  where $\rho(w)= ({\sqrt{1+4/\w^2}-2/\w})^{1/2}$.  This problem has the solution
\begin{equation}\label{eq:f_in}
    f(z) = \frac{1}{2\rho}\log(1+\rho z)-\frac{1}{2\rho}\log(1-\rho z),
\end{equation}
yielding the director angle 
\begin{equation}
    \theta(z) =- \arg f'(z)= \arg(1-\rho^2z^2).
\end{equation}
This solution hold for all anchoring strengths $w> 0$.

\subsection{Streamfunction}
Inserting the potential \eqref{eq:f_in} into \myeqref{eq:tpompsi}{b} yields the fluid pressure and vorticity,
\begin{equation}\label{eq:pom_in}
p-\im \omega=\A \left(g'(z) -\rho z\log\frac{1+\rho\bar{z}}{1-\rho\bar{z}}\right),
\end{equation}
while \myeqref{eq:tpompsi}{c} yields the streamfunction
\begin{equation}\label{eq:psi_in}
\psi=\frac{\A}{4} \Im\left[h(z)+\bar{z}g(z) -\frac{z^2}{2}\log(1-\rho^2\bar{z}^2)-\rho\frac{\bar{z}z^2}{2}\log\frac{1+\rho\bar{z}}{1-\rho\bar{z}}\right].
\end{equation}
The fluid velocity can  then be computed by taking a partial derivative of $\psi$ with respect to $z$, that is
\begin{equation}\label{eq:uv_in}
u-\im v=2\im \partial_z\psi=\frac{\A}{4}\left[h'(z)+\bar{z}g'(z)-\bar{g}(\bar{z})+\frac{\rho \bar{z}^2}{2}\log\frac{1-\rho z}{1+\rho z} -\rho\bar{z} z\log\frac{1+\rho \bar{z}}{1-\rho\bar{z}}-z\log(1-\rho^2 z^2)\right].
\end{equation}
Note that these expressions are  continuous in $|z|\leq1$ provided $h$ and $g$ are  holomorphic there. We now find the Goursat functions $h$ and $g$, which yield a fluid velocity, \eqref{eq:uv_in}, that vanishes on  $|z|=1$.

Consider the functions
\begin{subequations}\label{eq:Goursat_in}
\begin{align}
    g(z) &=\mathcal{G}(z)-\frac{1}{z}(1-\rho z)\log(1-\rho z )-\frac{1}{z}(1+\rho z)\log(1+\rho z)+\rho^2 z,\\
   \text{and} \quad   h(z)&=\mathcal{H}(z)+\frac{1}{2 z^2}(1-\rho z)\log(1-\rho z)+\frac{1}{2 z^2}(1+\rho z)\log(1+\rho z).
\end{align}
\end{subequations}
These are holomorphic in $|z|\leq1$ as long as $\mathcal{G}$ and $\mathcal{H}$ are holomorphic there. Furthermore, they yield   
\begin{equation}
  u-\im v= \frac{\A}{4 }\left[\mathcal{G}'(z)/z+\mathcal{H}'(z)-\bar{\mathcal{G}}(1/z)\right] \enspace \text{on}\enspace |z|=1,
  \end{equation}
which vanishes provided $\mathcal{H}'(z)=\bar{\mathcal{G}}(1/z)-(1/z)\mathcal{G}'(z)$.  As with the previous example, inserting a Taylor expansion for $\mathcal{G}(z)$ and $\mathcal{H}(z)$ around $z=0$ yields $\mathcal{G}(z)=a z+b$ and $\mathcal{H}(z)=\bar{b}z+c$ for $a\in\mathbb{R}$ and $b,c\in \mathbb{C}$. Here, $b$ and $c$ are arbitrary, and so  can be set to zero, while $a$  sets the reference pressure at the origin,
\begin{equation}
    p-\im \omega \sim \A a \enspace \text{as}\enspace |z|\to0,
\end{equation}
which we assume to vanish.

Overall, inserting \eqref{eq:Goursat_in} with $\mathcal{G}=\mathcal{H}=0$ into \eqref{eq:psi_in} yields the streamfunction given in the main text,
\begin{equation}
        \psi = \frac{\A }{8}(|z|^2-1)\Im\left[\frac{1-\rho z}{z^2}\log(1-\rho z)+\frac{1+\rho z}{z^2}\log(1+\rho z)\right],
\end{equation}
while the fluid pressure, vorticity, and velocity follow from \eqref{eq:pom_in} and \eqref{eq:uv_in}.  Furthermore, using $z_s=-\im z$, $\bar{z}=1/z$, and  $w =4\rho^2/(1-\rho^4)$, the  tractions \eqref{eq:app_trac_complex} acting on $|z|=1$ are given by
\begin{subequations}\label{eq:cylinder2_tractions}
\begin{align}
t^e_x-\im t^e_y&= \frac{\rho^2z(3-\rho^2 z^2)}{2(1-\rho^2 z^2)^2}+\frac{ z(z^2+\rho^2)}{2(z^2-\rho^2)^2}-\frac{1+3\rho^2}{1+\rho^2}\frac{1}{2z},\\
\text{and}\quad t^a_x-\im t^a_y&= \A\Er\left[z\log\left(1-\rho^2/z^2\right)+\frac{\rho }{ z^2}\log\frac{1-\rho z}{1+ \rho z}-\frac{1-\rho^4}{1/z-\rho^2 z}+\frac{2\rho^2}{ z}\right].
\end{align}
\end{subequations}

\subsection{Airy stress function}

The solid Goursat functions, $G(z)$ and $H(z)$,  must be locally holomorphic in $1<|z|<\infty$, yield a displacement that vanishes as $|z|\to\infty$, and satisfy the stress balance  \eqref{eq:app_stressbalance} on $|z|=1$. Integrating the surface tractions \eqref{eq:cylinder2_tractions} and using  $\bar{z}=1/z$ and $z_s=-\im z$, we find that the stress balance can be expressed as
\begin{equation}\label{eq:stressbalance2}
    \frac{1}{z}G'(z)-\overline{G}(1/z)+H'(z)=\int t^e_x-\im t^e_y\de s + \int t^a_x-\im t^a_y\de s,
\end{equation}
where
\begin{subequations}
\begin{align}
   \int t_x^e-\im t_y^e\de s &=-\frac{\rho \im}{2}\log\frac{1-\rho z}{1+\rho z}-\frac{\im z/2}{z^2-1/\rho^2}-\frac{\im z/2}{z^2-\rho^2}+\frac{1+3\rho^2}{1+\rho^2}\frac{\im}{2z},\\
  \text{and} \quad   \int t_x^a-\im t_y^a\de s &= \im \A\Er\left[z\log(1-\rho^2/z^2)+\rho\log\frac{z+\rho}{z-\rho}+\left(\frac{1}{2\rho}-\frac{\rho}{2 z^2}\right)\log\frac{1-\rho z}{1+\rho z }-\frac{\rho^2}{ z}\right].
\end{align}
\end{subequations}
Here, we can equate the $G'(z)$ and $H'(z)$ terms on the left-hand side of \eqref{eq:stressbalance2} with the terms holmorphic in $|z|>1$ on the right-hand side. Similarly,  the $\bar{G}(1/z)$ term is identified with the terms holomorphic in $|z|<1$.

Consider the functions
\begin{subequations}\label{eq:GH_2}
\begin{align}
    G(z) &= \mathcal{G}(z)+\frac{\im\rho}{2}\log\frac{z+\rho }{z-\rho}+\frac{\im \rho^2 z/2}{ z^2-\rho^2}+\im \A\Er\left[\frac{\rho^2z^2-1}{2\rho}\log\frac{z+\rho}{z-\rho}-\rho^2 z\right],\\
    \text{and} \quad  H(z) &=\mathcal{H}(z)+\frac{\im \rho^2}{1+\rho^2}\log z-\frac{\im\rho^2/2}{z^2-\rho^2}+\im \A\Er\left[\frac{z^2-1/\rho^2}{2}\log(1-\rho^2/z^2)+\rho^2\log z\right].
\end{align}
\end{subequations}
These are locally holomorphic in $1<|z|<\infty$ provided $\mathcal{G}(z)$ and $\mathcal{H}(z)$ are locally holomorphic there, also \eqref{eq:stressbalance2} is satisfied provided $\mathcal{H}'(z) = \bar{\mathcal{G}}(1/z)-\mathcal{G}'(z)/z$. Since $G(z)\sim\mathcal{G}(z)$ and $H'(z)\sim \mathcal{H}'(z)$ as $|z|\to\infty$, the displacement of the solid, \eqref{eq:displacements2}, satisfies
\begin{equation}
    2M(U-\im V)\sim \im\kappa\overline{\mathcal{G}(z)}+\im\bar{z}\mathcal{G}'(z)+\im \mathcal{H}'(z)\enspace \text{as}\enspace |z|\to\infty,
\end{equation}
which vanishes only if  $\mathcal{G}(z)=0$  and $\mathcal{H}(z)=c$ for some arbitrary constant $c\in\mathbb{C}$, which we set to zero.

Overall, inserting \eqref{eq:GH_2} with $\mathcal{G}=\mathcal{H}=0$ into \eqref{eq:Airystresses} and \eqref{eq:displacements2} yields the stress and  displacement of the solid. For large anchoring strengths, $\w\to \infty$, we find that
\begin{equation}
U+\im V\sim  \frac{w}{8M}\bigl[\mathcal{U}(z-1)-\mathcal{U}(-1-z)\bigr],
\end{equation}
for the  localized deformation 
\begin{equation}
\mathcal{U}(Z/w) =\kappa/(1+Z)+1/(1+\bar{Z})+(Z+\bar{Z})/(1+\bar{Z})^2.
\end{equation}
which is only apparent local to  the left and right poles, \ie~when $z\pm 1=\Oh(1/w)$. Note that this closely resembles the asymptotic expression found for the immersed cylinder example, as discussed in the main text. 

For a general anchoring strength, the aspect ratio of the deformed shape  is given by
\begin{equation}
 \mathcal{A}_R\sim 1+ U\bigr|_{z=1}-V\bigr|_{z=\im} = 1+(\mathcal{W}_1-\A\Er\mathcal{W}_2)/\M,
\end{equation}
with
\begin{subequations}
\begin{align}
    \mathcal{W}_1&=\frac{\kappa\rho}{2}\left(\arctanh \rho+\arctan\rho\right)+\frac{w}{8}(1+\kappa),\\
    \text{and}\quad     \mathcal{W}_2&=\frac{\kappa}{2\rho}(\arctanh \rho+\arctan\rho)+\rho(1-\kappa/2)(\arctanh\rho-\arctan\rho)+\frac{1}{2}\log(1-\rho^4).
\end{align}
\end{subequations}
Since $\mathcal{W}_1,\mathcal{W}_2\geq0$, we find that the deformed shape is elongated with the  $y$-axis if $\A\Er>\mathcal{W}_1/\mathcal{W}_2$ and elongated with the $x$-axis if $\A\Er<\mathcal{W}_1/\mathcal{W}_2$. Therefore, extensile activity ($\A<0$) always leads to parallel elongation, while contractile activity ($\A>0$) yields perpendicular elongation, as long as the activity is sufficiently strong.  Note that this is in  direct contrast to immersed cylinder example, which was discussed in the main text.

\bibliography{apssamp}